\documentclass[12pt,letterpaper]{article}
\usepackage{fullpage}

\usepackage{enumerate}
\usepackage{amsmath}
\usepackage{amsthm}
\usepackage{amssymb}
\usepackage{amscd}

\newtheorem{theorem}{Theorem}
\newtheorem{corollary}[theorem]{Corollary}
\newtheorem{proposition}[theorem]{Proposition}
\newtheorem{claim}[theorem]{Claim}
\newtheorem{lemma}[theorem]{Lemma}
\newtheorem{remark}[theorem]{Remark}
\newtheorem{Definition}[theorem]{Definition}
\newcommand{\E}{\textup{E}}
\renewcommand{\P}{\textup{P}}
\newcommand{\ignore}[1]{}
\def\eps{{\epsilon}}
\usepackage{color}

\definecolor{Red}{rgb}{1,0,0}
\definecolor{Blue}{rgb}{0,0,1}
\definecolor{Olive}{rgb}{0.41,0.55,0.13}
\definecolor{Green}{rgb}{0,1,0}
\definecolor{MGreen}{rgb}{0,0.8,0}
\definecolor{DGreen}{rgb}{0,0.55,0}
\definecolor{Yellow}{rgb}{1,1,0}
\definecolor{Cyan}{rgb}{0,1,1}
\definecolor{Magenta}{rgb}{1,0,1}
\definecolor{Orange}{rgb}{1,.5,0}
\definecolor{Violet}{rgb}{.5,0,.5}
\definecolor{Purple}{rgb}{.75,0,.25}
\definecolor{Brown}{rgb}{.75,.5,.25}
\definecolor{Grey}{rgb}{.5,.5,.5}
\definecolor{Black}{rgb}{0,0,0}
\definecolor{Black}{rgb}{0,0,0}

\def\a_{{}}

\newcommand{\ve}{\varepsilon}

\newcommand{\ga}{\gamma}
\newcommand{\la}{\lambda}
\newcommand{\al}{\alpha}
\newcommand{\be}{\beta}
\newcommand{\tiO}{\tilde{O}}
\newcommand{\ov}[1]{\overline{#1}}
\newcommand{\si}{\sigma}

\newcommand{\cD}{{\cal D}}
\newcommand{\cF}{{\cal F}}

\begin{document}

\title{Sorting from Noisy Information}
\author{Mark Braverman\thanks{Part of this work was done while the author was a graduate
student at the University of Toronto, supported by and NSERC CGS scholarship. Part of the work was done while the author was visiting IPAM, UCLA.} \\ Microsoft Research New England
\and Elchanan Mossel\thanks{Supported by an Alfred Sloan fellowship
in Mathematics, by NSF grants DMS-0528488
and DMS-0548249 (CAREER), by DOD ONR grant N0014-07-1-05-06 and by ISF grant 1300/08.
Part of this work was done while the author was visiting IPAM, UCLA.} \\Statistics and Computer Science, U.C. Berkeley and \\ Mathematics and CS, Weizmann Institute, Rehovot, Israel}

          \maketitle

\begin{abstract}
This paper studies problems of inferring order given noisy information.
In these problems there is an unknown order (permutation) $\pi$ on $n$ elements denoted by $1,\ldots,n$.
We assume that information is generated in a way correlated with $\pi$.
The goal is to find a maximum likelihood $\pi^*$ given the information observed.
We will consider two different types of observations: noisy comparisons and noisy orders.
\begin{itemize}
\item
{\em Noisy Orders (also called the Mallow's model)}.
Given the original permutation $\pi$, the probability of a permutation $\sigma$ being
generated is proportional to $e^{-\beta d_K(\sigma,\pi)}$. In other words, the probability is inverse exponential in the Kemeny distance of $\pi$ from $\sigma$, which
is the number of pairs ordered in $\pi$ differently from $\sigma$:
\begin{equation*}
d_K(\pi,\sigma) = \#\{ (i,j):~\pi(i)<\pi(j) \text{ and }\sigma(i)>\sigma(j)\}.
\end{equation*}
We assume that we are given $\sigma_1,\ldots,\sigma_r$ that are generated independently conditioned on $\pi$.
\item
{\em Noisy Comparisons}.
The input is the status of $\binom{n}{ 2}$
queries of the form $q(i,j)$, for $i < j$, where $q(i,j) = + (-)$
with probability $1/2+\la$  if $\pi(i) > \pi(j) (\pi(i) < \pi(j))$
for all pairs $i \neq j$, where $\la > 0$ is a constant. It is assumed
that the errors are independent. More generally, the input may be any collection of independent biased
signals on the order relationship between pairs of elements.
\end{itemize}

In this paper we present polynomial time algorithms for solving both problems with high probability.
For noisy orders the running time of the algorithm is $n^{1+O((\beta r)^{-1})}$, and for noisy comparisons the  algorithm runs in time
$n^{O(\la^{-3-\ve})}$.
Both algorithms have $O(n \log n)$ query complexity (with the constant depending on $\lambda,\beta$ and $r$).

As part of our proof we show that for both models
the maximum likelihood solution $\pi^{\ast}$ is close to the original permutation
$\pi$.
More formally, with high probability it holds that
\[
\sum_i |\pi(i) - \pi^{\ast}(i)| = \Theta(n),\quad
\max_i |\pi(i) - \pi^{\ast}(i)| = \Theta(\log n).
\]

Our results are of interest in applications to ranking, such as ranking in sports, or ranking of search items based on comparisons by experts.

\end{abstract}

\section{Introduction}
We study the problem of sorting in the presence of noise.
While sorting linear orders is a classical well studied problem,
the introduction of noise creates very interesting challenges.
Noise has to be considered when ranking or sorting is applied in many
real life scenarios.

A natural example comes from sports. How do we rank a league of soccer teams
based on the outcomes of the games? It is natural to assume that there is a true
underlying order of which team is better and that the game outcomes represent
noisy versions of the pairwise comparisons between teams. Note that in
this problem it is impossible to ``re-sample'' the order between
a pair of teams.
As a second example, consider experts ranking various items
according to their importance.
It is natural to assume that the experts' opinions
represent a noisy view of the actual order of significance. The question is
then how to aggregate this information?

\subsection{Aggregating  rankings: Mallow's Model}
The classical model for noisy permutations was introduced by Mallow~\cite{Mallow:57}.
This model is parameterized by a permutation $\pi^{\ast}$ and a real parameter $\beta > 0$.
The probability of observing a permutation $\pi$ is exponentially small in $\beta$ times the distance between
$\pi$ and $\pi^{\ast}$. More formally, given the
original permutation $\pi^*$, the probability of a permutation $\pi$ being
generated is inverse exponential in the Kemeny distance of $\pi$ from $\pi^*$. The Kemeny distance
is the number of pairs ordered in $\pi$ differently from $\pi^*$:
\begin{equation}
d_K(\pi,\pi^*) = \#\{ (i,j):~\pi^*(i)<\pi^*(j) \text{ and }\pi(i)>\pi(j)\}.
\end{equation}

\begin{Definition}
In Mallow's model, the probability of a permutation $\pi$ is given by
\begin{equation}
\label{eq:Mallow}
\P[\pi|\pi^*] = \frac{1}{Z(\be)} e^{-\be d_K(\pi,\pi^*)}.
\end{equation}
for a $\be>0$ and a normalization constant $Z(\be)$.
\end{Definition}
This model has been studied extensively in statistics and has been generalized in a number of ways,
see e.g.~\cite{Diaconis:88,FlingerVerducci:86,FlingerVerducci:88}.

Our goal is to find the best fit for the permutation $\pi^*$ given $r$ independent
observations $\pi_1,\ldots,\pi_r$ that are distributed according to \eqref{eq:Mallow}.
\begin{Definition}
The {\em Mallow Reconstruction Problem} (MRP) is the problem of finding
a $\pi^*$ maximizing the quantity
$$
\prod_{k=1}^r \Pi[\pi_k|\pi^*]=\frac{1}{Z(\be)^r} e^{-\be \sum_{k=1}^r d_K(\pi_k,\pi^*)},
$$
or equivalently minimizing
\begin{equation}
d(\pi^*) := \sum_{k=1}^r d_K(\pi_k,\pi^*).
\end{equation}
\end{Definition}
The optimization problem without any assumptions on the generating process is NP-hard~\cite{BaToTr:89}.
On the other hand, a number of heuristics were suggested in the statistical literature for solving the
problem~\cite{FlingerVerducci:90,CoScSi:99,MePhPaBl:07}. None of these heuristics have a guarantee to find the correct permutation even assuming the permutations are generated from the model.

In one of our main results we will show that the MRP problem can be solved in polynomial time, that approaches
linear time as $r$ increases.

\subsection{Aggregating noisy comparisons}
We next define a second model for noisy sorting. In this model the noise is applied to each pairwise comparison. In other words, for each pair, the correct order is observed with some probability greater than $1/2$.

\subsubsection{The sorting model: Noisy Signal Aggregation}
We will consider the following probabilistic model of instances.
There will be $n$ items denoted $1,\ldots,n$.
There will be a {\em true order} given by a permutation $\pi$ on $1,\ldots,n$. For two elements $i,j \in [n]$ we
write $i <_{\pi} j$ if $\pi(i) < \pi(j)$.

The algorithm will have access to $\binom{n}{2}$ signals defined as follows.

For each unordered pair $\{a,b\}$, it receives a signal $s_{a,b} = s_{b,a}$. The signal distribution $\cD$
depends on whether $a<b$ or $b<a$:
\begin{equation}
\cD = \left\{
\begin{array}{ll}
\cD_{a < b} & \text{if $\pi(a)<\pi(b)$},\\
\cD_{b < a} & \text{if $\pi(b)<\pi(a)$}.
\end{array}
\right.
\end{equation}
We assume that the signals are independent conditioned on the true
order. In other words, for any set $S=\{(i_1,j_1),(i_2,j_2),\ldots,(i_k,j_k)\}$
of unordered pairs, such that $(a,b)\notin S$, and a vector of signals $s= (s_{i_1,j_1}, s_{i_2,j_2},\ldots,
s_{i_k,j_k})$,
$$
\cD\left[s_{a,b}=\cdot~|~\pi, s\right] = \cD\left[s_{a,b}=\cdot~|~1_{\pi(a) < \pi(b)}\right].
$$

\noindent
The goal of {\em Noisy Signal Aggregation (NSA)}
problem defined below is to find a permutation $\pi$ that is most consistent with the signals.

\begin{Definition}
Given the signals $s_{i,j}$ for all pairs $\{i,j\}\in [n]$, the Noisy Signal Aggregation
is the maximum likelihood permutation $\pi$, assuming uniform prior. In other words,
$\pi$ maximizes the quantity
\begin{equation}
\P[\{s_{i,j}\}~|~\pi] = \prod_{i, j: i <_{\pi} j} \cD_{i < j}(s_{i,j}).
\end{equation}
\end{Definition}

\noindent
Given a signal $s_{a,b}$, assuming uniform prior we have
\begin{equation}
\frac{\P[a <_{\pi} b~|~s_{a,b}]}{\P[b <_{\pi} a~|~s_{a,b}]} =
\frac{\cD_{a < b}(s_{a,b})}{\cD_{b < a}(s_{a,b})}.
\end{equation}
We associate a score $q(a<b)$ with the decision to rank $a$ below $b$ as the log of this ratio:
\begin{equation}
q(a < b) :=
\log \frac{\cD_{a<b}(s_{a,b})}{\cD_{b<a}(s_{a,b})}.
\end{equation}
Obviously, $q(b<a) = -q(a<b)$.
Note that by Gibbs' inequality
$\E[q(a<b) | \pi(a) < \pi(b)]\ge 0$.
The NSA problem thus can be rephrased as the following problem.

\begin{proposition}
The NSA Problem is equivalent to the problem
of finding a $\sigma$ that
maximizes the total score
\begin{equation}
s_q (\sigma) := \sum_{i<_\sigma j} q(i<j).
\end{equation}
\end{proposition}

We will discuss several NSA models. The simplest one is defined as follows.
\begin{Definition}
The {\em Simple Noisy Sorting Aggregation}(SNSA) problems with parameter $\lambda$ is a NSA problem where
$s_{a,b} \in \{+,-\}$ for all $a,b$ and
\begin{equation} \label{eq:snsa}
\cD_{a>b}(+) = \frac{1}{2} + \lambda, \quad \cD_{a>b}(-) = \frac{1}{2} - \lambda,
\end{equation}
\[
\cD_{a<b}(+) = \frac{1}{2} - \lambda, \quad \cD_{a<b}(-) = \frac{1}{2} + \lambda.
\]
\end{Definition}
Our results  showing that the SNSA can be solved efficiently are presented in section~\ref{subsec:results}. The results are also extended to a much
more general family of NSA problems.

\subsubsection{Related Sorting Models and Results}~\label{subsec:related}
It is natural to consider the problem of finding a ranking $\sigma$ that
minimizes the score $s_q(\sigma)$ where the input $q$ takes only the values  of $\pm 1$  (a relation between every pair), and there are no probabilistic assumptions
on the input. This problem, called the {\em feedback arc set problem for tournaments}
is known to be NP hard~\cite{AiChNe:05,Alon:06}.
However, it does admit PTAS~\cite{KenyonSchudy:07} achieving a $(1+\eps)$
approximation for
\[
-\frac{1}{2}\left[s_q(\sigma)-{\binom{n}{2}}\right].
\]
in time that is polynomial in $n$ and doubly exponential in $1/\eps$.
The results of~\cite{KenyonSchudy:07} are the latest in a long line of work
starting in the 1960's and including~\cite{AiChNe:05,Alon:06}.
See~\cite{KenyonSchudy:07}
for a detailed history of the feedback arc set problem.

A problem that is in a sense easier than NSA is the problem where
repetitions are allowed in querying.
In this case it is easy to observe that the original
order may be recovered in $O(n \log^2 n)$ queries with high probability.
Indeed, one may perform any of the standard $O(n \log n)$ sorting
algorithms and repeat each query $O(\log n)$ times in order to obtain the
actual order between the queried elements with error probability $n^{-2}$ (say).
More sophisticated methods  show that in fact the true order may be
found in query complexity $O(n \log n)$ with high probability~\cite{FPRU:90},
see also~\cite{KarpKleinberg:07}.

\begin{remark}
Some of our results on the SNSA problem appeared as an extended abstract in~\cite{BravermanMossel:08}.
\end{remark}

\subsection{Main Results} \label{subsec:results}

\subsubsection{Mallow Reconstruction Problem}

For the Mallow Reconstruction Problem our main result is that the problem can be solved in time that tends
to linear as $r$ increases beyond $1/\be$. Formally, we prove the following:

\begin{theorem}
\label{thm:mrp-main}
There exists a randomized algorithm such that 
if $\pi_1,\ldots, \pi_r$ be rankings on $n$ elements independently generated by Mallow's model with parameter $\be>0$, and let $\al>0$.
Then a maximum probability order $\pi^m$ can is computed in time
$$
T(n)=O\left(n^{1+O\left(\frac{\al}{\be r}\right)} \cdot 2^{O\left(\frac{\al}{\be}+\frac{1}{\be^2}\right)}\cdot \log^2 n\right).
$$
and error probability $<n^{-\al}$. In particular, the algorithm tends to almost linear as  $r$ grows.
\end{theorem}

\subsubsection{Simple Noisy Signal Aggregation}
For the Simple Noisy Signal Aggregation problem, our main result is the following.
\begin{theorem} \label{thm:sort_simple}
For any $\la > 0$ and $\al > 0$
there exists a randomized algorithm that except with probability
at most $n^{-\al}$
finds an optimal solution to the Simple Noisy Signal Aggregation (SNSA)
with parameter $\la$ in time $n^{O((\al + 1) \la^{-4})}$.
\end{theorem}

\subsubsection{General Noisy Aggregation}
Our results extend to more general models of NSA aggregations which we now discuss.
In order for our aggregative reconstruction to work, we will need two properties from the signal distributions.

\begin{Definition}
\label{def:gabiased}
We say that a collection of distributions $\cD_{a<b},\cD_{b<a}$ is {\em strongly $\ga$-biased} if
\begin{enumerate}[(a)]
\item\label{gabiased:a}
For every ${\frac{ \log n }{ \ga} < m \le n}$, and for any $m$ different $\cD_{a_k,b_k}$ such that
$a_k <_{\pi} b_k$ for at least $2/3$ of the $k$'s:
\begin{equation}
\label{eq:gabiased}
\P\left[\sum_{k=1}^{m} q(a_k<b_k) > 0\right] > 1-2^{-\ga m}.
\end{equation}
\item \label{gabiased:b}
There is a constant $A$ such that for any $A$ different $\cD_{a_k,b_k}$ such that
$a_k < b_k$ holds for all the $k$'s,
\begin{equation}
\label{eq:gabiasedcondb}
\P\left[\sum_{k=1}^{A} q(a_k<b_k) > 0\right]>1-10^{-3}.
\end{equation}
\end{enumerate}
\end{Definition}

Under these conditions we prove the following. 
\begin{theorem} \label{thm:sort}
For any $\ga > 0$ and $\al > 0$
there exists a randomized algorithm that except with probability
at most $n^{-\al}$
finds an optimal solution to the Noisy Signal Aggregation (NSA)
problem on strongly $\ga$-biased signals in time $n^{\tiO((\al + 1) \gamma^{-3})}$.
\end{theorem}
In the statement above and throughout the paper $\tiO{}(\cdot)$ signifies order of magnitude up to logarithmic corrections in the variables in the expression inside the $\tiO{}(\cdot)$.
A key ingredient in the proof of Theorem~\ref{thm:sort} is the following.

\begin{theorem} \label{thm:dist}
Consider the NSA problem on strongly $\ga$-biased signals and let $\pi$ be the {\em true} order and
$\sigma$ be any optimal order. Let $\al > 0$. Then there exist constants
$c_1(\al,\ga)$ and $c_2(\al,\ga)$ such that except with probability $O(n^{-\al})$
the following inequalities hold:
\begin{equation} \label{eq:lin_dist}
\sum_{i=1}^n |\sigma(i) - \pi(i)| \leq c_1 n,
\end{equation}
\begin{equation} \label{eq:log_dev}
\max_i |\sigma(i) - \pi(i)| \leq c_2 \log n.
\end{equation}
\end{theorem}

Extending the techniques of~\cite{FPRU:90}
it is possible to obtain the results of
Theorem~\ref{thm:sort} with low sampling complexity. More formally,
\begin{theorem} \label{thm:sampling}
There is an implementation of a sorting algorithm with the same guarantees
as in Theorem~\ref{thm:sort} and whose sampling complexity is $C\, n \log n$
where $C = C(\al,\gamma,A)$.
\end{theorem}

In fact, Theorems \ref{thm:sort} and \ref{thm:dist} only require condition (\ref{gabiased:a}) from
Definition \ref{def:gabiased}. Condition (\ref{gabiased:b}) is only used to establish low sampling complexity
in Theorem \ref{thm:sampling}. We note that without condition (\ref{gabiased:b}) Theorem \ref{thm:sampling} holds
with sampling complexity of $O(n\log^2 n)$ rather than $O(n\log n)$.

We briefly note that from Azuma inequality it follows that
\begin{claim}
SNSA distributions~(\ref{eq:snsa}) with parameter $\la$ are strongly $\ga$ biased with $\gamma = \Omega(\lambda)$.
\end{claim}

Therefore Theorem~\ref{thm:sort_simple} follows from Theorems~\ref{thm:sort} and~\ref{thm:sampling}. More generally we have the following
claim that gives a large set of strongly $\ga$-biased distributions:

\begin{claim}
Consider the NSA problem where there exists a constant $C$ such that for all $a,b$ the functions $q(a,b)$ and $q(b,a)$
are bounded by $C$ and
$$
\E[q(a<b) | \pi(a) < \pi(b)] > \la
~~ \text{ and }
~~ \E[q(b<a) | \pi(b) < \pi(a)] > \la.
$$
Then the distributions $D_{a<b},D_{a>b}$ are strongly-$\gamma$ biased $\gamma = \Omega(\la/C)$.
\end{claim}






\subsection{Techniques}

\subsubsection{Mallow Reconstruction Problem}

In the Mallow Reconstruction Problem we need to aggregate $r$ noisy orderings $\pi_1, \ldots, \pi_r$ into
one optimal ordering $\pi^m$. It seems intuitively natural to try to ``average" these orderings into one
ordering $\pi$. It turns out that this intuition is correct, and in fact just taking the average of the locations
of element $x$ under the $\pi_i$'s locates it within a distance of $O(\frac{1}{\be r}\log n)$ from
its location in the true order $\pi^*$ with high probability. Note that this distance decreases as $r$ is increased.

Somewhat surprisingly, the bulk of the works goes into showing that the optimal ordering $\pi^m$ is pointwise
close to the true ordering $\pi^*$. This is important since we want to show that the ``average" $\pi$ is close
to $\pi^m$, but can only show that it is close to $\pi^*$.

Our algorithm uses the ``average" order $\pi$ as a starting point for
a dynamic programming algorithm from Section~\ref{sec:presorted} that finds the optimum $\pi^m$.
The results of this section may be of independent interest in cases where we are looking for an optimum order and have a
pointwise good initial guess for it.

\subsubsection{Noisy Signal Aggregation}

In order to obtain a polynomial time algorithm for the NSA problem it is important to
identify that any optimal solution to the problem is close to the true one.
Thus the main step of the analysis is the proof of Theorem~\ref{thm:dist}.

To perform the sorting efficiently we use an insertion algorithm. Given an optimal
order on a subset of the items we show how to insert a new element.
Since the optimal order both before and after the insertion of the element has
to satisfy Theorem~\ref{thm:dist}, it is also the case that no element moves
more than $O(\log n)$ after the insertion and re-sorting.
Using this we perform a ``re-sorting" using the dynamic programming algorithm
in Section~\ref{sec:presorted}.

The main task is proving Theorem~\ref{thm:dist} in
Section~\ref{sec:dist}. We first prove~(\ref{eq:lin_dist}) by showing that
for a large enough constant $c$, it
is unlikely that any order $\sigma$ whose total distance from the original order $\pi$
is more than $c n$ will have $s_q(\sigma) \geq s_q(\pi)$.
We then establish~(\ref{eq:log_dev}) in Section~\ref{subsec:log_dev}
using a bootstrap argument.
The argument is based on the idea that if the discrepancy in the position of
an element $a$ in an optimal order compared to the original order is more
than $c \log n$ for a large constant $c$, then there must exist many elements
that are ``close'' to $a$ that have also moved by much.
This then leads to a contradiction with~(\ref{eq:lin_dist}) applied to the neighborhood of $a$.

The final analysis of the insertion algorithm and the proof of Theorem~\ref{thm:sort} are provided in Section~\ref{sec:alg}. Section~\ref{sec:query} shows
how using a variant of the sorting algorithm it is possible to achieve
polynomial running time in sampling complexity $O_\ga(n \log n)$ thus
proving Theorem \ref{thm:sampling}.

\smallskip
\noindent
It is natural to ask whether the algorithm proposed here is applicable in the more
general feedback arc set problem and whether other efficient algorithms for the
more general problem are applicable here. It is easy to see that ``sorting by
number of wins'' algorithm, whose
approximation ratio has been recently studied \cite{CoFlRu:06},
 will result with high probability with an order
$\sigma'$ with $s_q(\sigma') > n^{3/2-\epsilon}+ s_q(\pi)$
for any $\eps > 0$ even for a simple Bernoulli $q$. A similar statement holds for a greedy algorithm
where elements are inserted optimally one at a time.
With more work it is possible to show that the algorithm presented here does
not provide a PTAS for the feedback arc set problem on tournaments and that
the complicated algorithm of~\cite{KenyonSchudy:07} does not solve
the problem presented here.

\subsubsection{Comparing the Two Sorting Problems}
It is interesting to compare the two sorting problems studied here.
The two generative models seem to be very closely related.
In fact it is easy to see that if one looks at the random tournament defined by the noisy
comparisons model and conditions on it being a permutation, then one recovers the Mallow model.
However, the conditioning on the tournament is a very strong conditioning as we condition on an event whose
probability is $2^{-\Omega(n^2)}$. This conditioning also has very strong consequences: for example -- with constant probability
the minimal element in the original $\pi^{\ast}$ will also be the minimal element in the generated order $\pi$.
Such a property does not hold for the noisy comparisons model as it is easy to see that the probability that the minimal element in
$\pi^{\ast}$ will satisfy the maximal number of less equal relations in the noisy input is $n^{-1/2 + o(1)}$.
In fact, as we will see below, in the noisy order model each generated permutation $\pi$ satisfies with high probability
that $\max | \pi^{\ast}(i) - \pi(i)| = O(\log n)$ so in a sense each permutation is already close to the original permutation.
For the noisy comparisons problem it is much harder to construct any permutation $\pi$ satisfying the condition above -- and this is one of the main algorithmic challenges we need to overcome.

\subsection{Distances between rankings}
Here we define a  measure of distance between rankings that will be used
later, and introduce some notation. First, given two permutations $\sigma$ and $\tau$ we define the
{\em dislocation distance} by
\[
d(\sigma,\tau) = \sum_{i=1}^n |\sigma(i) - \tau(i)|.
\]

Recall that the Kemeny distance $d_K(\si,\tau)$ is the number of pairs on which $\si$ and
$\tau$ disagree.
We will write $d(\sigma)$ for $d(\sigma,{id})$ where ${id}$ is the identity
permutation and $d_K(\si)$ for $d_K(\si,{id})$.
In this paper we will often use the following well known
claim~\cite{DiaconisGraham:77} relating the two distances.
\begin{claim}
\label{cl:sd}
For any $\tau$,
$$
\frac{1}{2} d(\tau) \le d_K(\tau) \le   d(\tau).
$$
\end{claim}



\subsection{Acknowledgment}
E. M. thanks Andrew Tomkins for inspirational discussions and Marina Meila for interesting discussions on Mallow's model.

\section{Sorting an almost sorted list} \label{sec:presorted}

In this section we present an algorithm that given a pre-sorted list so that each
element is at most $k$ positions away from its location in some optimal
ordering, finds an optimal ordering in time $O(n \cdot k^2 \cdot 2^{6 k})$.
The algorithm will be used as a building block for other algorithms in the paper.

\begin{lemma}
\label{lem:presort}
Let $[n]$ be $n$ elements
together with a scoring function $q$. Suppose that
we are given that there is an optimal ordering
$\si(1),\si(2),\ldots,\si(n)$, that maximizes the score
$$
s(\si) = \sum_{\si(i)<\si(j)} q(i<j),
$$
 such
that $|\si(i)-i|\le k$ for all $i$.
Then we can find such an optimal $\si$ in time $O(n\cdot k^2\cdot 2^{6k})$.
\end{lemma}
In the applications below $k$ will be $O(\log n)$.  When $k$ is small ($o(\log n)$), the algorithm tends to linear.
Note that a brute force search over all possible $\si$ would require time
$k^{\Theta(n)}$. Instead we use dynamic programming to reduce the running time.

\begin{proof}

We use a dynamic programming technique to find an optimal sorting.
Let $i<j$ be any
indices, then by the assumption, the elements in the optimally ordered interval
$$I=\{\a_{\si(i)}, \a_{\si(i+1)},\ldots, \a_{\si(j)}\}$$ satisfy
$I^{-} \subset I \subset I^{+}$ where
$$
I^+ = [i-k,j+k],\text{ and }
I^- = [i+k,j-k].
$$
Hence selecting the set $S_I=\{\si(i), \si(i+1),\ldots, \si(j)\}$ involves
choosing a set of size $j-i+1$ that contains the elements of $I^-$ and is contained in
$I^+$. This involves selecting $2k$ elements from the list (or from a subset of the list)
$$
[i-k,\ldots,i+k-1]\cup[j-k+1,\ldots,j+k]
$$
which has $4k$ elements. Thus the number of such $S_I$'s is bounded by $2^{4k}$.

We may assume without loss of generality that $n$ is an exact power of $2$.
Denote by
$I_0$ the interval containing all the elements. Denote by $I_1$ the left
half of $I_0$ and by $I_2$ its right half. Denote by $I_3$ the left half
of $I_1$ and so on. In total, we will have $n-1$ intervals of lengths $2,4,8,\ldots$.

For each $I_t=[i,\ldots,j]$ let $S_t$ denote the possible ($\le 2^{4k}$) sets of
the elements $I'_t=[\si(i),\ldots,\si(j)]$. We use dynamic programming to store an
optimal ordering $\si'$ of each such $I'_t \in S_t$. The total number of $I'_t$'s we will
have to consider is bounded by $n \cdot 2^{4k}$. In addition, for each processed interval
$I'_t$ we store its optimal score $s'(I'_t,\si')$, such that
$$
s'(I'_t,\si') = \sum_{\si'(i')<\si'(j'),~i'<j'<i'+2 k}  q(i'<j').
$$
In other words, we only sum over pairs $i', j'$ in $I_t'$ that are less than $2 k$ apart, and
which are the only pairs that potentially may get swapped. Note that the actual score $s(I'_t,\si')$ is shifted from
$s'(I'_t,\si')$ by an amount that is independent of $\si'$:
\begin{multline*}
s(I'_t,\si') = \sum_{\si'(i')<\si'(j')}  q(i'<j') =
\sum_{\si'(i')<\si'(j'),~i'<j'<i'+2 k}  q(i'<j') + \\ \sum_{\si'(i')<\si'(j'),~j'\ge i'+2 k} q(i'<j') =
s'(I'_t,\si') + \sum_{j'\ge i'+2 k}  q(i'<j').
\end{multline*}
Hence maximizing $s'(I'_t,\si')$ is equivalent to maximizing the actual score $s(I'_t,\si')$.

We proceed from $t=n-1$ down
to $t=0$ producing and storing an optimal sort for each possible $I'_t$. For
$t=n-1,n-2,\ldots,n/2$ the length of each $I'_t$ is $2$, and the optimal sort
can be found in $O(1)$ steps.

Now let $t<n/2$. We are trying to find an optimal sort of a given $I'_t=[i,i+2 s-1]$. We do
this by dividing the optimal sort into two halves $I_l$ and $I_r$ and trying to
sort them separately. We know that $I_l$ must contain all the elements in
$I'_t$ that come from the interval $[1,\ldots,{i+s-1-k}]$ and must be contained
in the interval $[{1},\ldots,{i+s-1+k}]$. Thus there are at most $2^{2k}$ choices for
the elements of $I_l$, and the choice of $I_l$ determines $I_r$ uniquely. For
each such choice we look up an optimum solution for $I_l$ and for $I_r$ in the
dynamic programming table.
Among all possible choices of $I_l$ we pick the best one. This is done
by recomputing the score $s'$ for the joined interval,
and takes at most $O(k^2)$ time, since the only new pairs $(i',j')$ with $|{i'}-{j'}|<2k$ are
along the boundary between $I_l$ and $I_r$. Thus the total
cost will be
\begin{multline*}
\sum_{i=1}^{\log n} \#\mbox{intervals of length $2^i$}\cdot \#\mbox{checks} \cdot \mbox{cost of check}  = \\
\sum_{i=1}^{\log n} O\left(\frac{n\cdot 2^{4k}}{2^i} \cdot 2^{2k}\cdot k^2 \right)=
O(n \cdot k^2 \cdot 2^{6k}).
\end{multline*}
\end{proof}

\section{Noisy ordering aggregation}

We will now turn our attention to aggregating noisy rankings generated by Mallow's model.
Recall that in this model, the probability of a permutation $\pi$ given a true ordering
$\pi^*$ is given by
\begin{equation}
\label{eq:malm}
\P[\pi|\pi^*]=\frac{1}{Z(\be)} e^{-\be d_K(\pi,\pi^*)},
\end{equation}
where $d_K(\pi,\pi^*)$ is the Kemeny distance -- the number of pairs  which $\pi$ and $\pi^*$ order differently.
As a first step we show that under this model, locations of individual elements are distributed
geometrically.

\begin{lemma}
\label{lem:mal1}
Let $a$ be an element that is ranked $k$-th by $\pi^*$. In other words, $\pi^*(a)=k$. Then
$$
\P[|\pi(a)-k|\ge i]< 2\cdot e^{-\be i}/(1-e^{-\be}).
$$
for all $i$.
\end{lemma}

\begin{proof}
For simplicity, we assume that $\pi^*$ is the identity map: $\pi^*(\a_i)=i$.
The key observation in the proof is that for any $m$, the distribution of
the locations of $\a_{m+1},\ldots, \a_n$ under $\pi$ remains the same if
we condition on the ordering of $\{\a_1,\ldots,\a_m\}$ between themselves
under $\pi$. Thus $\pi$ can be sampled by inserting the elements
$\a_1,\ldots,\a_n$ into the ordering one-by-one, each time conditioning
on the order so far.

Suppose we sampled the relative ordering of $\a_1,\ldots,\a_{k-1}$ under $\pi$, and
would like to insert a new element $\a_{k}$. By \eqref{eq:malm}, the probability of $\a_{k}$ being mapped to location
$k-i$ is bounded by $e^{-\be i}$. Note that after further insertions, the location of $\a_{k}$
may only increase. Hence
\begin{equation}
\label{eq:mal3}
\P[\pi(\a_k)\le k-i] < \sum_{j=i}^{\infty} e^{-\be j} = e^{-\be i}/(1-e^{-\be}).
\end{equation}
A symmetric argument gives the same bound for $\P[\pi(\a_k)\ge k+i]$, and completes the proof.
\end{proof}

Next, we assume that we are given $r$ independent samples generated by Mallow's model. In each
one of them, the location of $\a_k$ is geometrically distributed around $k$. This allows us to
prove a stronger concentration for the average of these locations. Again, for simplicity we
assume that $\pi^*$ is the identity $\pi^*(\a_i)=i$.

\begin{lemma}
\label{lem:mal2}
Suppose that the permutation $\pi_1,\ldots,\pi_r$ are drawn according to \eqref{eq:malm}. Let $a=\a_k$ be
the element ranked $k$-th by $\pi^*$. Let $\ov{\pi(a)}$ be the average index of $a$ under the permutations
$\pi_1,\ldots,\pi_r$:
$$
\ov{\pi(a)} = \frac{1}{r} \sum_{i=1}^r \pi_i(a).
$$
Then
$$
\P[|\ov{\pi(a)}-k|\ge i] \le 2 \cdot \left( \frac{(5 i+1) \cdot e^{-\be i}}{1-e^{-\be}} \right)^r
$$
for all $i$.
\end{lemma}

\begin{proof}
For a vector $b=(b_1,\ldots,b_r)$ of non-negative integers let $A_b$ denote the
event that $\pi_j(a)\le k-b_j$ for $j=1,\ldots,r$ for which $b_j>0$. By \eqref{eq:mal3} we have
$$
\P[A_b] < e^{-\be \sum_{j=1}^r b_j}/(1-e^{-\be})^r.
$$
Next, we note that the event $[\ov{\pi(a)}\le k -i]$ is covered by
$$
\bigcup_{\sum_{j=1}^r b_j = r\cdot  i} [A_b].
$$
Hence
\begin{multline*}
P[\ov{\pi(a)}\le k -i] < \#\left\{b: \sum_{j=1}^r b_j = r i \right\} \cdot \frac{e^{-\be r i}}{(1-e^{-\be})^r} =\\
{\binom{r i + r -1}{r-1}} \cdot \frac{e^{-\be r i}}{(1-e^{-\be})^r} < \frac{(5 i+1)^r \cdot e^{-\be r i}}{(1-e^{-\be})^r}.
\end{multline*}
Taking the symmetric bound for $P[\ov{\pi(a)}\ge k +i]$ completes the proof.
\end{proof}

In particular, assuming $r$ is fixed,   the following statement holds.

\begin{claim}
\label{cl:mal4}
Let $\al>0$. Then for sufficiently large $n$,
$$
\P\left[|\ov{\pi(\a_k)}-k|\ge \frac{\al+2}{\be \cdot r}\log n\text{ for some $k$}\right] < n^{-\al}.
$$
\end{claim}
\begin{proof}
The claim follows immediately from Lemma \ref{lem:mal2}.
\end{proof}

We see that the margin of error for each element decreases proportionally to $r$. We will now use Lemma \ref{lem:presort}
from Section \ref{sec:presorted} to give an efficient algorithm that finds the maximum likelihood permutation $\pi^{m}$ given
$\pi_1,\ldots,\pi_r$. Recall that such a $\pi^{m}$ minimizes
\begin{equation}
\label{eq:mal6}
\sum_{k=1}^r d_K(\pi_k,\pi^{m}) = \sum_{k=1}^r \sum_{\pi^{m}(\a_i)<\pi^{m}(\a_j)} 1_{ \pi_k(\a_i)>\pi_k(\a_j)} =
\sum_{\pi^{m}(\a_i)<\pi^{m}(\a_j)} \#\{k:\pi_k(\a_i)>\pi_k(\a_j)\}.
\end{equation}
Set $q(\a_i<\a_j):=\#\{k:\pi_k(\a_i)<\pi_k(\a_j)\}$. Then minimizing \eqref{eq:mal6} is equivalent to maximizing
$$
s(\pi^m) = \sum_{\pi^m(\a_i)<\pi^m(\a_j)} q(\a_i<\a_j).
$$

Let $\ov{\pi}$ be the elements $\{\a_k\}$ sorted according to their $\ov{\pi(\a_k)}$ value.
By Claim \ref{cl:mal4} it follows that except with probability $n^{-\al}$,
\begin{equation}\label{eq:mal7}
|\ov{\pi}(\a_k)-\pi^*(\a_k)| < 2 \cdot \frac{\al+2}{\be \cdot r}\log n \text{ for all $k$.}
\end{equation}

In order to apply Lemma \ref{lem:presort} to obtain the optimum $\pi^m$ from the approximation
$\ov{\pi}$ it remains to see that with high probability the optimum $\pi^m$ is pointwise close
to the original $\pi^*$ (and hence, by \eqref{eq:mal7}, to $\ov{\pi}$). For simplicity, we assume that
$\pi^*$ is the identity order $\a_1,\ldots,\a_n$.

Denote
$$
L =\max\left( 6 \cdot \frac{\al+2}{\be \cdot r}\log n,~ 6 \cdot \frac{\al+2+1/\be}{\be}\right).
$$
We first use \eqref{eq:mal3} to prove the following simple claim.

\begin{claim}
\label{cl:mal12}
Except with probability $n^{-\al}$ we have that for any $i$, $j$ such that
$i\le j-L$,
$$
q(\a_i<\a_j) > \frac{2}{3} r.
$$
\end{claim}

In other words, less than $1/3$ of the permutations $\pi_1,\ldots,\pi_r$ order $\a_i$ and $\a_j$ incorrectly.

\begin{proof}
By a direct application of \eqref{eq:mal3}, for each $k$,
\begin{multline*}
\P[\pi_k(\a_j)<\pi_k(\a_i)] \le \P[\pi_k(\a_j) \le j-L/2] +\P[\pi_k(\a_i) \ge i+L/2]  \le \\
2\cdot e^{-\be L/2}/(1-e^{-\be}) \le n^{-3 (\al+1)/r},
\end{multline*}
for a sufficiently large $n$. In the case when  $r\le \log n$, the probability of having at least
$r/3$ rearranged pairs is bounded by
$n^{-(\al+1)} \cdot 2^{r} < n^{-\al}$. In the case when $r>\log n$, we have
$$
\P[\pi_k(\a_j)<\pi_k(\a_i)] \le e^{-3(\al+1)},
$$
and the probability of having at least $r/3$ rearranged pairs is bounded by
$$
e^{-3(\al+1)\cdot r/3} \cdot 2^r < e^{-\al\cdot r}< n^{-\al}.
$$
\end{proof}

We are now ready to prove the lemma on the proximity of the optimum to the original.
\begin{lemma}
\label{lem:mal8}
Except with probability $<2\cdot n^{-\al}$, for any optimal $\pi^m$ and for all $k$, we have
$$
|\pi^m(\a_k)-\pi^*(\a_k)| \le 32 L,
$$
where $\pi^*$ is the original permutation.
\end{lemma}

\begin{proof}
We will assume that the sampled permutations $\pi_1, \ldots, \pi_r$ satisfy the property
in Claim \ref{cl:mal12}, which happens except with probability of at most $n^{-\al}$.
Suppose, for contradiction, that there is a $k$ such that $|\pi^m(\a_k)-k|=M>32 L$. Without loss
of generality suppose that $\pi^m(\a_k)=k+M$.

We first claim that there must be at least $T\ge M/4-L>7 L$ indexes $i<k$ such that $\pi^m(\a_i)\ge k$.
That is, many indexes move from below position $k$ to above position $k$.
Let $S$ be the set of indexes $j$ such that $k\le \pi^m(\a_j)<k+M$. We must have
$$
\sum_{j\in S} (q(\a_j < \a_k)-q(\a_j > \a_k)) > 0,
$$
for otherwise the permutation $\pi^{m}_0$ where $\a_k$ is moved back to location $k$ would score higher
than $\pi^m$. We spit $S$ into $S_1$, $S_2$ and $S_3$ as follows
$$
S=S_1\cup S_2\cup S_3 = \{j\in S:~j<k\}\cup \{j\in S:~k<j<k+L\}\cup\{j\in S:~j\ge k+L\}.
$$
Note that $|S_2|<L$. Hence, by our assumption,
\begin{multline*}
\sum_{j\in S} (q(\a_j < \a_k)-q(\a_j > \a_k))  =
\sum_{j\in S_1} (q(\a_j < \a_k)-q(\a_j > \a_k)) +\sum_{j\in S_2} (q(\a_j < \a_k)-q(\a_j > \a_k)) +\\ \sum_{j\in S_3} (q(\a_j < \a_k)-q(\a_j > \a_k)) <
r \cdot |S_1| +  r \cdot |S_2|  - ( r/3) \cdot |S_3| < r\cdot ( T+L) - (r/3) \cdot (M-T-L).
\end{multline*}
Hence $T+L - (M-T-L)/3 >0$, which implies that $T>7 L$.

The fact that there are $T$ indexes $i<k$ such that $\pi^m(\a_i)\ge k$, implies that there are at least $T$ indexes
$i\ge k$ with $\pi^m(\a_i)<k$. Denote
$$
T_1 = \{i<k: \pi^m(\a_i)\ge k\}, ~T_2 = \{i\ge k: \pi^m(\a_i)< k\}.
$$
Let $\pi^m_1$ be the permutation obtained from $\pi^m$ by concatenating its restriction to $H_L=\{\a_1,\ldots,\a_{k-1}\}$
with its restriction to $H_R=\{\a_k,\ldots, \a_n\}$. We claim that $\pi^m_1$, scores higher than $\pi^m$, which is a contradiction.
We first count the number of pairs $(\a_i<\a_j)$ on which $\pi^m$ and $\pi^m_1$ disagree such that $|i-j|<L$. To disagree, either
$\a_i$ or $\a_j$ has to belong to $T_1\cup T_2$, and in each case we have at most $L$ choices for the other. Hence the total number
of such pairs is at most $2 T L$. We denote these pairs by $P_1$.

Next we count the number of pairs $(\a_i<\a_j)$ on which $\pi^m$ and $\pi^m_1$ disagree such that $|i-j|\ge L$. Note that for
each such pair $\pi^m_1$ has the ``right" answer and we know that in this case $q(\a_i<\a_j)>(2/3) r$. Each of the elements of
$T_1$ participates in such a pair with each element of $T_2$, save at most $L$ elements for which $|i-j|<L$. Thus
the number of such pairs is at least $T(T-L)$. We denote them by $P_2$.

The final difference in score between $\pi^m$ and $\pi^m_1$ is given by
\begin{multline*}
s(\pi^m_1)-s(\pi^m)=\sum_{(\a_i<\a_j)\in P_1} (q(\a_i<\a_j)-q(\a_j<\a_i))+ \sum_{(\a_i<\a_j)\in P_2} (q(\a_i<\a_j)-q(\a_j<\a_i)) >\\
(-r) \cdot |P_1| + (r/3) \cdot |P_2| \ge (-r) (2 T L) + (r/3) (T^2- T L) = r(T^2/3 - 7 T L /3)>0,
\end{multline*}
since $T>7 L$. Contradiction.
\end{proof}

It follows from Lemma~\ref{lem:mal8} and Claim~\ref{cl:mal4} that the pointwise distance between
$\ov{\pi}$ and $\pi^m$ is bounded by $k =33 L$.  We can now apply Lemma \ref{lem:presort}  to obtain:

\medskip \noindent
{\bf Theorem \ref{thm:mrp-main}.~}
Let $\pi_1,\ldots, \pi_r$ be rankings on $n$ elements independently generated by Mallow's model with parameter $\be>0$, and let $\al>0$.
Then a maximum probability order $\pi^m$ can be computed in time
$$
T(n)=O\left(n^{1+O\left(\frac{\al}{\be r}\right)} \cdot 2^{O\left(\frac{\al}{\be}+\frac{1}{\be^2}\right)}\cdot \log^2 n\right).
$$
except with probability $<n^{-\al}$. In particular, the algorithm tends to almost linear as  $r$ grows.

\smallskip
\noindent
{\bf Remark.} It should be noted that since the $\pi_i$'s are actual orderings, they can be recovered
with $O(n \log n)$ queries of the type $\a_j <^?_{\pi_i} \a_k$ each. Thus the total query complexity is
trivially bounded by $O(r n \log n)$.

\section{Noisy comparisons aggregation}

\subsection{The Discrepancy between the true order and optimal orders}
\label{sec:dist}

The goal of this section is to establish that with high probability any optimum
solution will not be far from the original solution. We first establish that
the orders are close on average, and then that they are pointwise close to each
other.

\subsubsection{Average proximity} \label{subsec:lin_dist}

We prove that with high probability, the total difference between the original and any optimal ordering is linear in the length of the interval.

We begin by bounding the probability that a specific permutation $\sigma$
will beat the original ordering. Recall that $d_K(\si)$ is the number of pairs on
which the permutation $\si$ disagrees with the identity.

\begin{lemma} \label{lem:two_perms} Assume that the distributions of the
scoring functions are strongly $\ga$-biased, and
suppose that the original ordering is $\a_1 < \a_2 \ldots < \a_n$.
Let $\sigma$ be another permutation. Then the probability that
$\sigma$ beats the identity permutation is bounded from above by
\[
2^{-\ga d_K({\sigma})}.
\]
\end{lemma}

\begin{proof}
In order for $\sigma$ to beat the identity, it needs to beat it in the $d_K({\sigma})$ positions
 where they differ. The probability bound follows immediately from the definition
 of  $\ga$-biased distributions.
\end{proof}

\noindent
Recall that $d(\tau)=\sum_{i=1}^n |\tau(i)-i|$ is the total dislocation of elements
under $\tau$.

\begin{lemma} \label{lem:distance_distribution}
The number of permutations $\tau$ on $[n]$ satisfying $d(\tau) \leq c\,n$
is at most
\[
  2^n\, 2^{(1+c)\, n \,H(1/(1+c))}.
\]
\end{lemma}

Here $H(x)$ is the binary entropy of $x$ defined by
$$
H(x) = - x\log_2 x - (1-x) \log_2(1-x) < -2 x \log_2 x,
$$
for small $x$.

\begin{proof}
Note
that each $\tau$ can be uniquely specified by the values of
$s(i)=\tau(i)-i$, and that we are given that $\sum |s(i)|$ is exactly
$d(\tau) \le c n$. Thus there is an injection of $\tau$'s with $d(\tau)=m$ into
sequences of $n$ numbers which in absolute values add up to $m$.
It thus suffices to bound the number of such sequences.
The number of unsigned sequences equals the number of
ways of placing $m$ balls in $n$ bins, which is equal to $\binom{n+m-1}{n-1}$.
Signs multiply the possibilities by at most $2^n$. Hence the total number of
$\tau$'s with $d(\tau)=m$ is bounded by $2^n \cdot {\binom{n+m-1}{n-1}}$. Summing
up over the possible values of $m$ we obtain
$$
\sum_{m=0}^{c n} 2^n \cdot {\binom{n+m-1} {n-1}} < 2^n \cdot {\binom{n+ c n}{n}} \\
\leq 2^n\, 2^{(n+ c n)\,H(n/(n+c n))}.
$$
\end{proof}

\begin{lemma}\label{lem:p0}
Suppose that the true ordering is $\a_1 < \ldots < \a_n$ and $n$ is large enough.
Then if $c \geq 1$ and
\[
\ga c >4\cdot( 1 + (1+c) H(1/(1+c))),
\]
the probability that any ranking $\sigma$ is optimal and
$d(\sigma) > c n$ is at most $2^{-c n \ga/5}$ for sufficiently large $n$.
In particular, as $\ga \to 0$, it suffices to take
\[
c = O(\ga^{-1} \log 1/\ga) = \tiO (\ga^{-1}).
\]
\end{lemma}

\begin{proof}
Let $\sigma$ be an ordering with $d(\sigma) > c n$. Then by Claim~\ref{cl:sd}
we have $d_K({\sigma}) > c n /2$. Therefore the probability that such
an ordering will beat the identity is bounded by
$2^{-c n \ga/2}$ by Lemma~\ref{lem:two_perms}.
We now use union bound and Lemma~\ref{lem:distance_distribution} to obtain
the desired result.
\end{proof}

\subsubsection{Pointwise proximity} \label{subsec:log_dev}

In the previous section we have seen that it is unlikely that the {\em average} element
in the optimal order is more than a constant number of positions away from its original
location. Our next goal is to show that the {\em maximum} dislocation of an element
is bounded by $O(\log n)$.
As a first step, we show that one ``big" dislocation
is likely to entail many ``big" dislocations.

\begin{lemma}
\label{lem:p1}
Suppose that the true ordering of $\a_1,\ldots,\a_n$ is given by the identity ranking, that is, $\a_1 < \a_2 \ldots < \a_n$. Let $1 \leq i < j \leq n$
be two indices and $m=j-i$.
Let $A_{ij}$ be the event that there is an optimum ordering $\si$ such
that $\si(i)=j$ and the following two conditions hold:
\[
[i,j] \subset \si[i- 2 m,j+2 m],
\]
\[
\left|\left(\si[1,i-\ell-1] \cup \si[j+\ell+1,n]\right) \cap [i,j-1]\right| \leq \ell,
\]
i.e.,  elements from at most $2m$-away are mapped to $[i,j]$ by $\si$, and
at most $\ell$ elements are mapped to the interval $[i,j-1]$ from
outside the interval $[i-\ell,j+\ell]$ by $\si$. We set
 $\ell=\left\lfloor c_\ell m\right\rfloor<m$, where $$c_\ell = \frac{\ga}{300 (1-\log \ga)} = \tilde{\Omega}(\ga).$$ Then
$$
P(A_{ij})<2^{- m\ga/2}.
$$
\end{lemma}

\begin{proof}
We prove the lemma by applying a union bound over all possible variants
of the set $B=\si^{-1}[i,j]$. We know that $B$ may contain a subset of
size at most $3 \ell$ of elements coming from $[i-m,i-1]\cup[j+1,j+m]$, thus
the number of possible sets is bounded by
$$
{\binom{5m} {3\ell}} \cdot {\binom{m} {3\ell}} \le 2^{5 m \cdot H\left( \frac{3\ell}{5 m}
\right) + m \cdot H\left( \frac{3\ell}{ m}
\right)} <
2^{6 m \cdot H\left( \frac{3\ell}{5 m}
\right)} < 2^{12 m \cdot \frac{3 \ell}{5  m} \cdot \log \frac{5 m}{3 \ell}} < 2^{m\ga/2}.
$$

The assumption that $\si$ is optimal implies in particular that moving the
$i$-th element from the $j$-th position where it is mapped by $\si$ back to the
$i$-th position does not improve the solution. For each specific choice of $B$, more than
$2/3$ of the elements that are mapped to $[i,j-1]$ are originally smaller than $\a_i$, and
hence the probability of moving the $i$-th element back not improving the solution
is bounded by $2^{-m\ga}$. By union bound,
$$
P[A_{ij}] < 2^{m\ga/2} \cdot 2^{-m\ga} = 2^{-m \ga/2}.
$$
\end{proof}

As a corollary to Lemma \ref{lem:p1} we obtain the following using a simple
union-bound. For the rest of the proof all the $\log$'s are base $2$.

\begin{corollary}
\label{cor:p1}
Let
$$
m_1=(-\log \ve +2\log n)/(\ga/2) = O((-\log \ve+ \log n)/ \ga),
$$
 then $A_{ij}$ does not occur for any $i,j$ with
$|i-j|\ge m_1$ with probability $>1-\ve$.
\end{corollary}

Next, we formulate a corollary to Lemma \ref{lem:p0}.

\begin{corollary}
\label{cor:p0}
 Suppose that $\a_1<\a_2<\ldots<\a_n$ is
the true ordering. Set $$m_2=2 m_1.$$ For each interval $I=[\a_i,\ldots,\a_j]$ with
at least $m_2$ elements consider
all the sets $S_I$  which   contain the elements from
$$
I^- = [\a_{i+m_2},\ldots,\a_{j-m_2}],
$$
and are contained in the interval
$$
I^+ = [\a_{i-m_2},\ldots,\a_{j+m_2}].
$$
Then with probability $>1-\ve$ all such sets $S_I$ do not have an optimal ordering that has
a total deviation
from the true of more than $c_2\, |i-j|$, with
$$c_2 = \frac{35}{\ga} = O(\ga^{-1}),$$
 a constant.
\end{corollary}

\begin{proof}
There are at most $n^2 \cdot 2^{4 m_2}$ such sets. The probability of each set
not satisfying the conclusion is bounded by Lemma \ref{lem:p0} with
$$
2^{-c_2 m_2 \ga/5} =   2^{-7 m_2} =
 2^{-m_2}\cdot 2^{-2 m_2} \cdot 2^{-4 m_2} <
\ve \cdot n^{-2} \cdot 2^{-4 m_2}.
$$
The last inequality holds because $m_2>\max(\log n,-\log \ve)$.
By taking a union bound over all the sets we obtain the statement of the corollary.
\end{proof}

We are now ready to prove the main result on the pointwise distance between an optimal
ordering and the original.

\begin{lemma}
\label{lem:pmain}
Assuming that the events from Corollaries \ref{cor:p1} and \ref{cor:p0} hold,
it follows that for each optimal ordering $\si$ and for each $i$, $|i-\si(i)|< c_3\log n$,
where
$$c_3 = \frac{24}{c_\ell^2} \cdot \frac{m_2}{\log n} = \tiO(\ga^{-3}(-\log \ve/\log n + 1))$$
 is a constant.
 In particular, this conclusion holds with probability $>1-2 \ve$.
\end{lemma}

\begin{proof}
 We say that a position $i$ is {\em good} if
there is no index $j$ such that $\si(j)$ is on the other side of $i$ from $j$ and
$|\si(j)-j|\ge m_2$. In other words, $i$ is good if there is
no ''long'' jump over $i$ in $\si$.
In the case when $i=j$ or $i=\si(j)$ for a long
jump, it is not considered good. An index that is not good is bad.
An interval $I$ is
bad if all of its indices are bad. Our goal is to show that there are no bad intervals of length $\ge c_3 \log n$.
This would prove the lemma, since if there is an $i$ with
$|i-\sigma(i)| >   c_3 \log n$ then there is a bad interval of length at least
$c_3 \log n$.

Assume, for contradiction, that $I=[i,\ldots,{i+t-1}]$ is a bad interval
of length $t\ge c_3 \log n$, such that $i-1$ and $i+t$ are both good (or lie
beyond the endpoints of $[1,\ldots,n]$). Denote by $S$ the set of elements
that is mapped to $I$ by $\si$. Denote the  indices in $S$ in their original
order by $i_1<i_2<\ldots<i_t$, i.e., we have:
$\{\sigma(i_1),\ldots,\sigma(i_t)\} = I$.

By the goodness of the endpoints of $I$ we have
$$
[i+m_2, i+t-1-m_2] \subset \{ i_1,\ldots,i_t \} \\ \subset [i-m_2, i+t-1+m_2].
$$
Denote the permutation induced by $\si$ on $S$ by $\si'$ so
$\sigma(i_j) < \sigma(i_{j'})$ is equivalent to $\si'(j) < \si'(j')$.
 The permutation
$\si'$ is optimal, for otherwise it would have been possible to improve $\si$
by improving $\si'$.

By Corollary \ref{cor:p0} and Claim \ref{cl:sd}, we have the following bound on
the number of switches under $\si'$ (and hence the number of switches
on the elements of $S$ between themselves under $\si$):
\[
d_K({\si'})\le d(\si')\le c_2 t.
\]
In how many switches can
the elements of $S$ participate under $\si$? They participate in switches with other
elements of $S$ to a total of $d_K({\si'})$. In addition, they participate in switches
with elements that are not in $S$. These elements must originate at the margins of the
interval $I$: either in the interval $[i-m_2,i+m_2]$ or the interval $[i+t-1-m_2,i+t-1+m_2]$.
Thus, each contributes at most $2 m_2$ switches with elements of $S$. There are at most $2 m_2$
such elements. Hence the total number of switches between elements in $S$ and in $\ov{S}$ is at most
$4 m_2^2$. Hence
\begin{equation}
\label{eq1}
\sum_{i\in S} |\si(i)-i| \le \sum_{i\in S} \#\{\mbox{switches $i$ participates in}\}  \le
4 m_2^2 + 2 d_K({\si'}) \le 4 m_2^2 + 2 c_2 t.
\end{equation}

We assumed that the entire interval $I$ is bad, hence for every position $i$ there is
an index $j_i$ such that $|\si(j_i)-j_i|\ge m_2$ and such that $i$ is in the interval
$J_i=[j_i,\si(j_i)]$ (or the interval $[\si(j_i),j_i]$, depending on the order).
Consider all such $J_i$'s. We will say that an interval $J_i$ is {\em free}
if there is no interval $J_j$ intersecting it such that $|J_j|> 2|J_i|$.
We will use a Vitali covering lemma argument to show that we can choose
a disjoint collection of free intervals whose total length is at least $|I|/5$.

Let $\cF$ be the collection of $J_i$'s that are free. We claim that for every $i\in I$
there is an element $J_i \in \cF$ such that the ``tripling" of $J_i$: $J_i^3=[j_i-|J_i|,\si(j_i)+|J_i|]$ covers $i$.
We know that there is an interval $J_1$ that covers $i$. If $J_1$ is free, then we are done.
Otherwise, there is an interval $J_2$ that intersects $J_1$ and is at least twice as long.
We continue this process until we reach an interval $J_k$ that is free. How far can $i$ be
from the endpoints of $J_k$? At most
$$
|J_{k-1}|+|J_{k-2}|+\ldots+|J_1| < |J_k|.
$$
Thus, the tripling of $J_k$ covers $i$.

The argument now proceeds as follows: Order the intervals  in $\cF$
in a decreasing length order (break ties arbitrarily). Go through the list and
add a $J_i$ to our collection if it is disjoint from all the currently selected intervals.
We obtain a collection $J_1, \ldots, J_k$ of disjoint intervals of the form $[j_i,\si(j_i)]$.
Denote the length of the $i$-th interval by $t_i = |j_i-\si(j_i)|\ge m_2$.
Let $J_i^5$ be the ''quintupling" of the interval $J_i$: $J_i^5=[j_i-2 t_i,\si(j_i)+2 t_i]$.
We claim that the $J_i^5$-s cover the entire interval $I$. Let $m$ be a position on the interval
$I$. Then there is an interval $J$ in $\cF$ such that its tripling $J^3$ covers
$m$. Choose the longest such interval $J'=[j,\si(j)]$. If $J'$ has been selected to
our collection then we are done. If not, it means that $J'$ intersects a longer interval
$J_i$ that has been selected. This means that the tripling of $J'$ is covered by the quintupled interval $J_i^5$.
In particular, $m$ is covered by $J_i^5$. We conclude that
$$
t= \mbox{length}(I) \le \sum_{i=1}^{k} \mbox{length}(J_i^5) = 5 \sum_{i=1}^{k} t_i.
$$
Thus $\sum_{i=1}^{k} t_i \ge t/5$. This concludes the covering argument.

We now apply Corollary \ref{cor:p1} to the intervals $J_i$. Since every $J_i$ is
free, we conclude that on an interval
$J_i$ the contribution of the elements of $S$ that are mapped to $J_i$ to the sum of
deviations under $\si$ is at least $\ell_i^2$ where $\ell_i = c_\ell t_i$.
Thus
\begin{multline*}
\sum_{i\in S} |\si(i)-i| \ge \sum_{j=1}^{k} \ell_j^2  = c_\ell^2 \cdot \sum_{j=1}^{k} t_j^2
\ge c_\ell^2 \cdot m_2 \cdot \sum_{j=1}^{k} t_j
\ge c_\ell^2 \cdot m_2 \cdot t/5  \\ \ge m_2 \cdot \frac{c_\ell^2}{6}  \cdot c_3 \log n +
\frac{c_\ell^2}{30}  \cdot m_2 t
> m_2 \cdot (4 m_2)+ 2 c_2 t =4 m_2^2 + 2 c_2 t,
\end{multline*}
for sufficiently large $n$.
The result contradicts \eqref{eq1} above.
Hence there are no bad intervals of length $\ge c_3 \log n$, which completes the proof.
\end{proof}


\subsection{The algorithm} \label{sec:alg}

We are now ready to give an algorithm for computing the optimal ordering with high
probability in polynomial time. Note that Lemma \ref{lem:pmain} holds for any interval
of length $\le n$ (not just length exactly $n$). Set $\ve = n^{-\al-1}/4$.
Given an input, let
$S\subset \{\a_1,\ldots,\a_n\}$ be a random set of size $k$. The probability that there is an
optimal ordering $\si$ of $S$ and an index $i$ such that $|i-\si(i)|\ge c_3 \log n$, where
$$
c_3 = \tiO(\ga^{-3} (-\log \ve/\log n +1)) =\tiO(\ga^{-3} (\al +1)),
$$
is bounded by $2\ve$ by  Lemma \ref{lem:pmain}. Let
$$S_1\subset S_2\subset\ldots\subset S_n$$
be a randomly selected chain of sets such that $|S_k|=k$. Then the probability that
an element of an optimal order of any of the  $S_k$'s deviates from its original
location by more than $c_3 \log n$    is bounded by
$2 n \ve =  n^{-\al}/2$. We obtain:

\begin{lemma}
\label{lem:chain}
Let $S_1\subset\ldots\subset S_n$ be a chain of randomly chosen subsets with $|S_k|=k$.
Denote by $\si_k$ an optimal ordering on $S_k$.
Then with probability $\ge 1-n^{-\al}/2$,
for each  $\si_k$ and for each $i$, $|i-\si_k(i)|<c_3\log n$, where
$c_3 = \tiO(\ga^{-3}(\al+1))$ is a constant.
\end{lemma}

We are now ready to prove the main result, Theorem~\ref{thm:sort}, which we restate

\begin{theorem}
\label{thm:main}
There is an algorithm that runs in time $n^{c_4}$, where
$$c_4= \tiO(\ga^{-3}(\al+1))$$ is a constant,
that outputs an optimal ordering with probability $\ge 1-n^{-\al}$.
\end{theorem}

\begin{proof}
First, we choose a random chain of sets $S_1\subset\ldots\subset S_n$ such
that $|S_k|=k$. Then by Lemma \ref{lem:chain}, with probability $1- n^{-\al}/2$,
 for each optimal order $\si_k$ of $S_k$ and for each $i$, $|i-\si_k(i)|<c_3\log n$.
 We will find the orders $\si_k$ iteratively until we reach $\si_n$ which will be
 an optimal order for our problem. Denote $\{a_k\}=S_k-S_{k-1}$. Suppose that we have
 computed $\si_{k-1}$ and we would like to compute $\si_k$. We first  insert
 $a_k$ into a location that is close to its original location as follows.

Recall that $c_3 = \tilde{\Theta}(\ga^{-3}(\al+1)) > (\al+3)/\ga$.
Break $S_k$ into blocks $B_1, B_2,\ldots, B_s$ of length $c_3 \log n$. We claim
that with probability $>n^{-\al-1}/2$ we can pinpoint the block $a_k$ belongs to
within an error of $\pm 2$, thus locating $a_k$ within $3 c_3 \log n$ of its original
location.

Suppose that $a_k$ should belong to block $B_i$. Then by our assumption on
$\si_{k-1}$, $a_k$ is bigger than any element in $B_1,\ldots,B_{i-2}$ and smaller
than any element in $B_{i+2},\ldots,B_s$. By comparing $a_k$ to each element in
the block and taking the sum of the comparison scores, we see that the probability of having an incorrect
comparison result with a block $B_j$ is bounded by $n^{-\al-2}/2$. Hence
the probability that $a_k$ will not be placed correctly up to an error of two
blocks is bounded by $n^{-\al-1}/2$ using union bound.


  Hence after inserting $a_k$ we obtain an ordering of $S_k$ in which each element is at most
  $3 c_3 \log n$ positions away from its original location. Hence each element is at most $4 c_3 \log n$
  positions away from its optimal location in $\si_k$. Thus, by Lemma \ref{lem:presort} we can obtain
  $\si_k$ in time $O(n^{24 c_3+2})$. The process is then repeated.

  The probability of each stage failing is bounded by $n^{-\al-1}/2$. Hence the probability of the
  algorithm failing assuming the chain  $S_1\subset\ldots\subset S_n$ satisfies Lemma \ref{lem:chain}
  is bounded by $n^{-\al}/2$.
  Thus the algorithm runs in time $O(n^{24 c_3 +3})=n^{\tiO(\ga^{-3}(\al+1))}$ and has a failure probability of
  at most
 $  n^{-\al}/2 +  n^{-\al}/2 =  n^{-\al}.$
\end{proof}

\subsection{Query Complexity} \label{sec:query}

In this section we outline the proof of Theorem~\ref{thm:sampling}. Recall that the
theorem states that
although the running time of the algorithm is a polynomial of $n$ whose degree
depends on $\ga$, the query complexity of a variant of the algorithm is
$O(n \log n)$. In this section we demonstrate that our algorithm can be implemented with high probability
 using only $O(n\log n)$ queries.
Note that there are two types of queries in the algorithm. The first type
is comparing elements in the dynamic programming, while the second is when inserting new elements.
We will show that both parts require only $O(n\log n)$ queries. We start with queries in
the dynamic programming part.

\begin{lemma} \label{lem:dynamic_query}
For all $\al > 0, \gamma < 1/2$ there exists $c(\al,\gamma) < \infty$
such that the total number of comparisons performed in the dynamic
programming stage  of the algorithm is at most
$c\, n \log n$ except with probability $O(n^{-\al}/4)$.
\end{lemma}

\begin{proof}
Recall that in the dynamic programming stage, each element is compared with
elements that are at current distance at most $c_3 \log n$ from it, where
$c_3 = c_3(\al,\gamma)=\tilde{O}(\ga^{-3}(\al+1))$.

Consider a random insertion order of the elements $a_1,\ldots,a_n$.
Let $S_{n/2}$ denote the set of elements inserted up to the $n/2$-th insertion.
Then by standard concentration results it follows that there exists
$c_5(c_3,\al)$ such that for all $1 \leq i \leq n - c_5 \log n$ it holds that
\begin{equation} \label{eq:up_int_query}
|[a_i,a_i+c_5 \log n] \cap S_{n/2}| \geq c_3 \log n,
\end{equation}
and for all $c_5 \log n \leq i \leq n$ it holds that
\begin{equation} \label{eq:low_int_query}
|[a_i-c_5 \log n,a_i] \cap S_{n/2}| \geq c_3 \log n
\end{equation}
except with probability at most $n^{-\al-1}$.
Note that when~(\ref{eq:up_int_query}) and~(\ref{eq:low_int_query}) both hold
the number of different queries used in the dynamic programming while inserting the
elements from $\{ a_1,\ldots,a_n \} \setminus S_{n/2}$ is at most $2 c_5 n \log n$, since none
of these elements is ever compared to an element that is further than $c_5\log n$ away
from it in the true order.

Repeating the argument above for the insertions performed from
$S_{n/4}$ to $S_{n/2}$, from $S_{n/8}$ to $S_{n/4}$ etc. we obtain that the total
number of queries used is bounded by:
\[
2 c_5 \log n (n + n/2 + \ldots + 1) \leq 4 c_5 n \log n,
\]
except with probability $< n^{-\al}/4$. This concludes the proof.
\end{proof}

Next we show that there is implementation of insertion that requires only
$O(\log n)$ comparisons per insertion. To this end, we recall condition (\ref{gabiased:b}) from
Definition \ref{def:gabiased} of strongly $\ga$-biased distributions.

\begin{itemize}
\item[(b)] There is a constant $A$ such that for any $A$ different $\cD_{a_k,b_k}$ such that
$a_k < b_k$ holds for all the $k$'s,
\begin{equation}
\label{eq:gabiasedcondbrepeat}
\P\left[\sum_{k=1}^{A} q(a_k<b_k) > 0\right]>1-10^{-3}.
\end{equation}
\end{itemize}


\begin{lemma} \label{lem:insertion_query}
For all $\alpha > 0$, $A \geq 1$ and $\gamma > 0$ there exists a
$$C(A,\gamma, \al)=\tilde{O}((A \ +\ga^{-3})(\al+1))$$
such that except with probability $O(n^{-\al-2}/2)$ it is
 possible to perform the insertion in the proof of Theorem \ref{thm:main}
so that each element is inserted using at most $C \log n$
comparisons, $O(\log n)$ time
and the element
is placed a distance of at most $4 c_3 \log n$ from its optimal location, as required by the algorithm.
\end{lemma}

\begin{proof}
Bellow we maintain the notation that $c_3(\al,\gamma)=\tilde{O}(\ga^{-3}(\al+1))$
is such that at all stages
of the insertion and for each item, the distance between the location of the
item in the original order and the optimal order is at most $c_3 \log n$.
This will result in an error with probability at most $n^{-\al}/2$.



Let $c_6=O(\al+1)$ be chosen so that
\begin{equation} \label{eq:rw_bias}
\P\left[Bin(c_6 \log n, 0.99) < \frac{c_6}{2} \log n + 2 \log_2 n\right] < n^{-\al-3},
\end{equation}
Let $c_7 = A c_6 + 4 c_3$.

We now describe an insertion step.
Let $S$ denote a currently optimally sorted set. We will partition $S$ into
consecutive intervals of length between $c_7 \log n$ and $2 c_7 \log n$ denoted
$I_1,\ldots,I_{t}$. We will use the notation $I_i'$
for the sub-interval of $I_i = [s,t]$ defined by
$I_i' = [s+2 c_3 \log n,t-2 c_3 \log n]$.
We say that a newly inserted element $j$ {\em belongs}
to one of the interval $I_i$
if one of the two closest elements to it in the original order belongs to
$I_i$. Note that $j$ can belong to at most two intervals. An element in $S$ belongs to $I_i$ iff it is one of the elements in $I_i$.
Note furthermore
that if $j$ belongs to the interval $I_i$ then its optimal insertion location
is determined up to $2 (A c_6 + 6 c_3) \log n$. Similarly, if we know
it belongs to one of two intervals then its optimal insertion location
is determined up to $$c_8 \log n := 4 (A c_6 + 6 c_3) \log n.$$

Note that by the choice of $c_3$
we may assume that all elements belonging to $I_i$ are
smaller than all elements of $I_j'$ if $i < j$ in the true order.
Similarly, all elements belonging to $I_j$ are larger than all elements of
$I_j'$ if $j > i$.
We define formally the interval $I_0 = I_0'$ to be an interval of elements that are smaller than all the items and the interval $I_{t+1} = I_{t+1}'$
to be an interval of elements that is bigger than all items.

We construct a binary search tree on the set $[1,t]$ labeled by
sub-intervals of $[1,t]$ such that the root is labeled by $[1,t]$ and
if a node is labeled by an interval $[s_1,s_2]$ with
$s_2 - s_1 > 1$ then its two children
are labeled by $[s_1,s']$ and $[s',s_2]$, where $s'$ is chosen so that the
length of the two intervals is the same up to $\pm 1$. Note that the two
sub-interval overlap at $s'$. This branching process terminates at intervals
of the form $[s,s+1]$. Each such node will have a path of descendants of length
$c_6 \log n$ all labeled by $[s,s+1]$.

We use a variant of binary search described in Section 3 of~\cite{FPRU:90}. The algorithm
will run for $c_6 \log n$ steps starting at the root of the tree.
At each step the algorithm will proceed from a node of the tree to either
one of the two children of the node or to the parent of that node.

Suppose that the algorithm is at the node labeled by $[s_1,s_2]$ and
$s_2 - s_1 > 1$.
The algorithm will first take $A$ elements from $I_{s_1-1}'$ that have not been
explored before and will check that the current item is greater than the
majority of them. Similarly, it will make a comparison with $A$ elements from
$I_{s_2+1}'$. If either test fails it would backtrack to the
parent of the current node. Note that if the test fails then it is the case
that the element does not belong to $[s_1,s_2]$ except with probability
$<0.01$.

Otherwise, let $[s_1,s']$ and $[s',s_2]$ denote the two children of
$[s_1,s_2]$. The algorithm will now perform a majority test against $A$ elements from $I_{s'}$ according to which it would choose one of the
two sub-intervals $[s_1,s']$ or $[s',s_2]$. Note again that a correct
sub-interval is chosen except with probability at most $0.01$ (note that
in this case there may be two ``correct'' intervals).

In the case where $s_2 = s_1+1$ we perform only the first test. If it fails
we move to the parent of the node. It it succeeds, we move to the single child.
Again, note that we will move toward the leaf if the interval is correct
with probability at least $0.99$. Similarly, we will move away from the leaf
if the interval is incorrect with probability at least $0.99$.

Overall, the analysis shows that at each step we move toward a leaf including
the correct interval with probability at least $0.99$. From~\eqref{eq:rw_bias} it follows that with probability at least $1-n^{-\al-3}$ after $c_6 \log n$ steps
the label of the current node will be $[s,s+1]$ where the inserted element
belongs to either $I_s$ or $I_{s+1}$. Thus the total number
of queries is bounded by $3 A c_6 \log n$.

Now, once we have located the element within $c_8 \log n$ positions, we can refine the search by comparing the
element to the relevant blocks $B_j$ from the algorithm in Theorem \ref{thm:main}. Thus will take at most
$c_8 \log n$ more queries, to a grand total of
$$
c_8 \log n + 3 A c_6 \log n  = \tilde{O}((A \ +\ga^{-3})(\al+1))
$$
queries to execute the insertion step of the algorithm.
This concluded the proof.
\end{proof}

\bibliographystyle{alpha}
\bibliography{my,all}

\begin{thebibliography}{MPPB07}

\bibitem[ACN05]{AiChNe:05}
N.~Ailon, M.~Charikar, and A.~Newman.
\newblock Aggregating inconsistent information: ranking and clustering.
\newblock In {\em Proceedings of 37th STOC}, 2005.

\bibitem[Alo06]{Alon:06}
N.~Alon.
\newblock Ranking tournaments.
\newblock {\em Siam Journal on Discrete Mathematics}, 20(1):137--142, 2006.

\bibitem[BM08]{BravermanMossel:08}
M.~Braverman and E.~Mossel.
\newblock Noisy sorting without resampling.
\newblock In {\em Proceedings of the nineteenth annual ACM-SIAM symposium on
  Discrete algorithms (SODA)}, page 268, 2008.

\bibitem[BTT89]{BaToTr:89}
J.~Bartholdi, III, C.~A. Tovey, and M.~A. Trick.
\newblock Voting schemes for which it can be difficult to tell who won the
  election.
\newblock {\em Soc. Choice Welf.}, 6(2):157--165, 1989.

\bibitem[CFR06]{CoFlRu:06}
Don Coppersmith, Lisa Fleischer, and Atri Rudra.
\newblock Ordering by weighted number of wins gives a good ranking for weighted
  tournaments.
\newblock In {\em SODA '06: Proceedings of the seventeenth annual ACM-SIAM
  symposium on Discrete algorithm}, pages 776--782, New York, NY, USA, 2006.
  ACM.

\bibitem[CSS99]{CoScSi:99}
William~W. Cohen, Robert~E. Schapire, and Yoram Singer.
\newblock Learning to order things.
\newblock {\em J. Artificial Intelligence Res.}, 10:243--270 (electronic),
  1999.

\bibitem[DG77]{DiaconisGraham:77}
Persi Diaconis and R.~L. Graham.
\newblock Spearman's footrule as a measure of disarray.
\newblock {\em J. Roy. Statist. Soc. Ser. B}, 39(2):262--268, 1977.

\bibitem[Dia88]{Diaconis:88}
Persi Diaconis.
\newblock {\em Group representations in probability and statistics}.
\newblock Institute of Mathematical Statistics Lecture Notes---Monograph
  Series, 11. Institute of Mathematical Statistics, Hayward, CA, 1988.

\bibitem[FPRU90]{FPRU:90}
U.~Feige, D.~Peleg, P.~Raghavan, and E.~Upfal.
\newblock Computing with unreliable information.
\newblock In {\em Proceedings 22nd STOC}, 1990.

\bibitem[FV86]{FlingerVerducci:86}
M.~A. Flinger and J.S. Verducci.
\newblock Distance based ranking models.
\newblock {\em Journal of the Royal Statistical Society B}, 48:359--369, 1986.

\bibitem[FV88]{FlingerVerducci:88}
M.~A. Flinger and J.S. Verducci.
\newblock Multistage ranking models.
\newblock {\em J. Amer. Statist. Assoc.}, 83(403):892--901, 1988.

\bibitem[FV90]{FlingerVerducci:90}
M.~A. Flinger and J.S. Verducci.
\newblock Posterior probability for a consensus ordering.
\newblock {\em Psychometrika}, 55:53--63, 1990.

\bibitem[KK07]{KarpKleinberg:07}
D.~Karp and B.~Kleinberg.
\newblock Noisy binary serach and its applications.
\newblock In {\em Proceedings of 11th SODA}, pages 891--890, 2007.

\bibitem[KMS07]{KenyonSchudy:07}
C.~Kenyon-Mathieu and W.~Schudy.
\newblock How to rank with few errors.
\newblock In {\em Proceedings of 39th STOC}, pages 95--103, 2007.

\bibitem[Mal57]{Mallow:57}
C.~L. Mallows.
\newblock Non-null ranking models.
\newblock {\em Biometrika}, 44:114--130, 1957.

\bibitem[MPPB07]{MePhPaBl:07}
M.~Meila, K.~Phandis, A.~Patterson, and J.~Blimes.
\newblock Consenus ranking under the exponential model.
\newblock Preprint, 2007.

\end{thebibliography}

\end{document}